\documentclass[conference]{IEEEtran}
\IEEEoverridecommandlockouts
\usepackage{cite}
\usepackage{amsmath,amssymb,amsfonts}
\usepackage{algorithmic}
\usepackage{graphicx}
\usepackage{textcomp}
\usepackage{xcolor}
\usepackage{quantikz}
\usepackage{caption}
\usepackage{subcaption}
\usepackage{flushend}
\usepackage{enumerate}
\usepackage{hyperref}
\usepackage{bbm}
\usepackage{pgfplots}
\def\BibTeX{{\rm B\kern-.05em{\sc i\kern-.025em b}\kern-.08em
    T\kern-.1667em\lower.7ex\hbox{E}\kern-.125emX}}

\DeclareCaptionLabelSeparator{periodspace}{.\quad}

\begin{document}
\bstctlcite{IEEEexample:BSTcontrol}

\title{Quantum Algorithms for tensor-SVD}

\author{\IEEEauthorblockN{Jezer Jojo\IEEEauthorrefmark{1}, Ankit Khandelwal\IEEEauthorrefmark{2} and M Girish Chandra\IEEEauthorrefmark{2}}
\IEEEauthorblockA{\IEEEauthorrefmark{1}Indian Institute of Science Education and Research, Pune, India}
\IEEEauthorblockA{\IEEEauthorrefmark{2}TCS Research, Tata Consultancy Services, India\\
Email: jezer.vallivattam@students.iiserpune.ac.in, khandelwal.ankit3@tcs.com, m.gchandra@tcs.com}}

\maketitle

\begin{abstract}

A promising area of applications for quantum computing is in linear algebra problems.
In this work, we introduce two new quantum t-SVD (tensor-SVD) algorithms. The first algorithm is largely based on previous work that proposed a quantum t-SVD algorithm for context-aware recommendation systems. The new algorithm however seeks to address and fix certain drawbacks to the original, and is fundamentally different in its approach compared to the existing work.
The second algorithm proposed uses a hybrid variational approach largely based on a known variational quantum SVD algorithm.
\end{abstract}

\begin{IEEEkeywords}
quantum computing, quantum algorithms, quantum machine learning, linear algebra
\end{IEEEkeywords}

\section{Introduction}

The search for quantum algorithms that may outperform classical computing is a matter of much interest. Given the inherent linearity of quantum mechanics, a substantial amount of work is being done attempting to solve linear algebra problems using quantum computers.

The Quantum Recommendation System \cite{kp} was one such attempt. It is a quantum algorithm that takes a matrix with limited entries of users' preferences for certain products, and uses a low-rank matrix approximation to extrapolate the missing entries. This algorithm has a runtime that is polylogarithmic in the size of the matrix.

Taking inspiration from this algorithm, however, Tang  \cite{tang} proposed a classical algorithm that is also polylogarithmic in the size of the preference matrix, effectively `dequantizing' the quantum algorithm.

In \cite{tsvd}, Wang et al. introduce a quantum algorithm for Context-Aware Recommender Systems that uses the t-SVD (tensor-SVD) to obtain a low-rank approximation of a `preference tensor' that contains entries of users' preferences towards products in different `contexts'. This algorithm makes use of the routines introduced in \cite{kp}, but is harder to `dequantize'. Similarly in \cite{qhosvd}, Gu et al. introduce a quantum HOSVD (Higher Order Singular Value Decomposition) algorithm that provides an exponential speedup over the classical HOSVD algorithm. 

The t-SVD \cite{tsvddef} is one of many generalizations of the Singular Value Decomposition (SVD) for third-order tensors. It has various applications including video denoising \cite{tsvddenoise}, context-aware recommendation systems, and image classification \cite{tsvdimg}.

Compared to other tensor generalizations of SVD such as the CP decomposition \cite{cp} or Tucker decompositions \cite{tucker} like HOSVD, one of the things that makes a quantum approach to t-SVD promising \cite{tsvd} is that the t-SVD involves computations in the Fourier space, and applying a Fourier Transform is something quantum computers are inherently good at.


In this work, we first identify drawbacks in the algorithm from \cite{tsvd} and propose a new quantum algorithm that addresses these drawbacks. Our algorithm when applied to context-aware recommendation systems is polylogarithmic in the size of the tensor. Then we introduce a new hybrid variational algorithm that can also be used to obtain the t-SVD of a given tensor.

In Section \ref{sec:tSVDdef}, we introduce our notation and define what a t-SVD is in a simple way that suits our purposes.

Then in Section \ref{sec:TrtSVD}, we describe our first quantum t-SVD algorithm. This is a purely quantum (as opposed to a hybrid algorithm) largely motivated by \cite{tsvd}. In Section \ref{sec:VtSVD}, we then describe our hybrid quantum t-SVD algorithm which takes inspiration from the variational quantum SVD introduced in \cite{wang}.

Finally, we discuss our results and future outlooks in Section \ref{sec:disc}.

\section{Notation}\label{sec:tSVDdef}
\subsection{Tensor notation and t-SVD definition}
Consider an order-3 tensor $\mathcal{A}\in \mathbb{R}^{M\times N\times L}$ with elements  $a_{ijk}$; $i<M,j<N,k<L$.\\
$A^{(k)}$ or $\mathcal{A}[:,:,k]$ will denote the $k$-th face of $\mathcal{A}$ which is defined as the matrix whose $(i,j)$-th element is $a_{ijk}$.\\
Similarly we can define the $(i,j)$-th tube of $\mathcal{A}$, denoted by $\mathcal{A}[i,j,:]$, as the vector whose $k$-th entry is $a_{ijk}$.\\
Let $\tilde{\mathcal{A}}\in \mathbb{R}^{M\times N\times L}$ be the tensor whose $(i,j)$-th tube $\tilde{\mathcal{A}}[i,j,:]$ is the Fourier transform of $\mathcal{A}[i,j,:]$. The tilde denotes that a Fourier transform is applied along this third dimension. The $k$-th face of $\tilde{\mathcal{A}}$ is denoted by $\tilde{A}^{(k)}$ or $\tilde{\mathcal{A}}[:,:,k]$.

To find the t-SVD of an order-3 tensor $\mathcal{A}\in\mathbb{R}^{M\times N\times L}$ is to find order-3 tensors $\mathcal{U}\in\mathbb{R}^{M\times M\times L}$, $\mathcal{S}\in\mathbb{R}^{M\times N\times L}$, and $\mathcal{V}\in\mathbb{R}^{N\times N\times L}$ such that the SVD of $\tilde{A}^{(k)}$ is $\tilde{U}^{(k)}\tilde{S}^{(k)}\tilde{V}^{(k)\dagger}$ for all $0\leq k<L$ \cite{tsvddef}.

In this paper, we will choose to refer to the entries of $\tilde{\mathcal{S}}$ as the singular values of $\mathcal{A}$, denoted as
\begin{equation}
    \tilde{\sigma}_i^{(k)}:=\tilde{\mathcal{S}}[i,i,k].
\end{equation}

We will also define $\tilde{u_i}^{(k)}$ and $\tilde{v_i}^{(k)}$ as follows:

\begin{align}
    \tilde{u_i}^{(k)}:=&\tilde{\mathcal{U}}[:,i,k]\\
    \tilde{v_i}^{(k)}:=&\tilde{\mathcal{V}}[:,i,k]
\end{align}

\subsection{Quantum computing notation}
All kets represent normalized quantum states. Unless otherwise mentioned, the following conventions are used:

\begin{itemize}
    \item If $x$ is a positive integer, $|x\rangle$ represents the computational basis state corresponding to the bitstring representation of $x$.
    \item If $x$ can be negative, \textit{two's complement} representation is used.
    \item $|\tilde{x}\rangle$ represents $QFT|x\rangle$, where $QFT$ is the Quantum Fourier Transform operator.
    \item If $w$ is a vector, $|w\rangle$ represents the state whose statevector is $w/\|w\|$.
    \item If $M$ is a matrix or tensor, $|M\rangle$ represents the state whose statevector is the flattened vector $M_{flat}/\|M\|_F$.
    \item We use $|\dots\rangle$ to represent a general normalized state when we don't want to explicitly label or describe the specific state. We use this for error terms or states we will end up throwing away through postselection. 
\end{itemize}

\section{Purely quantum algorithm for tensor completion using \normalfont{t}-SVD}\label{sec:TrtSVD}

In this section, we first explore the limitations of an existing quantum t-SVD algorithm, thus indicating the opportunity for an improved solution. We then go through a procedure that serves the same purpose as the Quantum Singular Value Estimation (Quantum SVE) from \cite{kp} but adapted for the t-SVD. Lastly, we will see how this routine can be used for tensor completion and then for recommendation systems.

\subsection{Drawbacks of existing algorithm}
Reference \cite{tsvd} makes use of the Quantum SVE routine from \cite{kp} to construct a quantum context-aware recommendation algorithm based on the t-SVD. That work however uses a \textit{linear combination of unitaries} \cite{lcu} approach to create a block encoding of the required walk operator that they then need to perform Quantum Phase Estimation on (Appendix B of \cite{tsvd}). This step is nontrivial and its most straightforward application would require an auxiliary register of qubits whose size grows exponentially with the number of phase register qubits (Lemma 4 (Product of block-encoded matrices) in \cite{powerblock}).
It also makes strong implicit assumptions of the tensor in question. Appendix B of \cite{tsvd} defines $U_{P_k}$ and $U_{\hat{P}_m}$ as follows:

\begin{align}
    U_{P_k}|i\rangle|0\rangle&=\frac{1}{\|\mathcal{A}(i,:,k)\|_2}\sum_{j=0}^{N-1}a_{ijk}|i\rangle|j\rangle\\
    U_{\hat{P}_m}&\overset{\Delta}{=}\frac{1}{L}\sum^{L-1}_{k=0}\omega^{km}U_{P_{k}}
\end{align}

It is then asserted that

\begin{equation}
    U_{\hat{P}_m}|i\rangle|0\rangle = |i\rangle|\Tilde{\mathcal{A}}[i,:,m]\rangle.
\end{equation}

For a given $i$ and up to a scaling factor, this can only hold under the assumption that the norm $\|\mathcal{A}(i,:,k)\|_2$ is the same for all values of $k$.

The quantum algorithm proposed in this section does not contain these drawbacks. It is fundamentally different in its procedure and works by modifying the Quantum SVE routine itself.\\

\subsection{Quantum t-SVE}
As proposed in \cite{kp}, the Quantum SVE with respect to a matrix $A$ is used to extract the singular value corresponding to a given amplitude encoded state of the left singular vector of $A$. Similarly, the Quantum t-SVE algorithm proposed here, when applied with respect to a tensor $\mathcal{A}$, can be used to extract the singular value $\tilde{\sigma}_i^{(k)}$ from the corresponding statevector $|\tilde{v_i}^{(k)}\rangle|\tilde{k}\rangle$.

\subsubsection{Circuit registers}
This procedure makes use of $6$ registers:
\begin{enumerate}[(i)]
    \item Walk operator left singular vector register $(1)$, with $n_m:=\lceil \log_2{(M)}\rceil$ qubits
    \item Right singular vector register $(2)$, with $n_n:=\lceil \log_2{(N)}\rceil$ qubits
    \item Walk operator face register $(3)$, with $n_l:=\lceil \log_2{(L)}\rceil$ qubits
    \item Face register $(4)$, with $n_l$ qubits
    \item Phase register $(phase)$, with $n_p$ qubits
    \item Abs Phase register $(abs)$, with $d$ qubits
\end{enumerate}

\subsubsection{Data structure requirements}\label{sec:datastruct}
As input to this procedure we need access to operators $P'$ and $Q'$ such that

\begin{align}
    &P'_{1234}|i\rangle_1|0\rangle_2|\tilde{k}\rangle_3|0\rangle_4\nonumber\\
    =&\frac{1}{\|\mathcal{A}\|_F}\sum_j\|\mathcal{A}[:,j,:]\|_F|i\rangle_1|j\rangle_2|\tilde{k}\rangle_3|\tilde{k}\rangle_4\\
    &Q'_{1234}|0\rangle_1|j\rangle_2|0\rangle_3|\tilde{l}\rangle_4\nonumber\\
    =&\frac{1}{\|\mathcal{A}[:,j,:]\|_F}\sum_{i,k}a_{ijk}|i\rangle_1|j\rangle_2|k\rangle_3|\tilde{l}\rangle_4.
\end{align}

$P'$ can of course be further decomposed as an operator $P''$ acting on register $2$, and an operation copying register $3$ to register $4$ in the Fourier basis. $P''$ is defined such that

\begin{equation}
    P''|0\rangle=\frac{1}{\|\mathcal{A}\|_F}\sum_j\|\mathcal{A}[:,j,:]\|_F|j\rangle.
\end{equation}

The same data structure proposed in \cite{kp} can be used to store a matrix whose $(j,i\times 2^{n_l}+k)$-th entry is $\mathcal{A}[i,j,k]$ and, with such a data structure, the operators $P''$ and $Q'$ can be applied with an associated circuit depth that's $\text{polylog}(LMN)$. The circuit depth associated with copying register $3$ to register $4$ in the Fourier basis is $O(n_l)$ which means the circuit depth spent in applying $P'$ is also $\text{polylog}(LMN)$.

\subsubsection{Algorithm Input-Output and Complexity}\label{sec:qtsvedefs}
Let $\mathcal{A}$ be the given tensor stored in the data structure described in Section \ref{sec:datastruct}. Define $\theta_i^{(k)}$ as follows:

\begin{equation}
    \theta_i^{(k)}=2\arccos{\frac{\tilde{\sigma}_i^{(k)}}{\|\mathcal{A}\|_F}}
\end{equation}

Let $\delta>0$ be the smallest difference between $\theta$ values.

\begin{equation}
    \delta:=\min_{i,j,k,l}{(|\theta_i^{(k)}-\theta_j^{(l)}|)}
\end{equation}

We pick $d$ to be an integer such that

\begin{equation}
    2^{-d}<\frac{\delta}{2\pi}.
\end{equation}

Let $\epsilon>0$ be the precision factor.

Then, we can construct a circuit $QtSVE$ such that

\begin{align}
    &QtSVE|\tilde{v_i}^{(k)}\rangle_2|\tilde{k}\rangle_4\nonumber\\
    =&|\tilde{v_i}^{(k)}\rangle_2|\tilde{k}\rangle_4|\theta_i^{(k)}\rangle_{abs}+\sqrt{2}\overline{\epsilon}|\dots\rangle.\label{eq:binom}
\end{align}

Here, $|\theta_i^{(k)}\rangle$ represents a state whose first $d$ qubits form a basis state approximation of $2^d\times\theta_i^{(k)}/(2\pi)$. Here, $\overline{\epsilon}$ represents any value less than or equal to $\epsilon$.


While we could employ quantum arithmetic to further compute $|\tilde{\sigma}_i^{(k)}\rangle$, it is not a necessary step for the algorithms discussed in this paper. And if we wanted to use this routine directly to compute and measure the singular values of a tensor, it would be more practical to measure $|\theta_i^{(k)}\rangle$ and then use it to compute $\tilde{\sigma}_i^{(k)}$ classically.

The time complexity of this procedure is given by

\begin{equation}
    \text{Circuit depth}=O\left(\frac{\text{polylog}(LMN)}{\delta\epsilon^2}\right).
\end{equation}

\subsubsection{Theoretical underpinnings}

Our aim would be to construct operators $2PP^\dagger - \mathbbm{1}$ and $2QQ^\dagger - \mathbbm{1}$ for a $P$ and $Q$ that satisfy the following properties

\begin{align}
    P^\dagger P &= \mathbbm{1}\\
    Q^\dagger Q &= \mathbbm{1}\\
    P^\dagger Q &= \sum_kA^{(k)} |\tilde{k}\rangle\langle \tilde{k}|
\end{align}
where $|\tilde{k}\rangle=QFT|k\rangle$.\\

To do this, we define matrices $P$ and $Q$ like so

\begin{align}
    P=&\frac{1}{\|\mathcal{A}\|_F}\sum_{i,j,k}\|\mathcal{A}[:,j,:]\|_F(|i\rangle|j\rangle|\tilde{k}\rangle|\tilde{k}\rangle)(\langle i|\langle\tilde{k}|)\\
    Q=&\frac{1}{\|\mathcal{A}[:,j,:]\|_F}\sum_{i,j,k}a_{ijk}(|i\rangle|j\rangle|k\rangle|\tilde{l}\rangle)(\langle j|\langle \tilde{l}|)
\end{align}

To implement $W=(2PP^\dagger - \mathbbm{1})(2QQ^\dagger - \mathbbm{1})$, we can use our operations $P'$ and $Q'$.


\begin{equation}
    2PP^\dagger-\mathbbm{1}=P'_{1234}(2|0\rangle_2|0\rangle_4\langle 0|_2\langle 0|_4-\mathbbm{1})P'^\dagger_{1234}
\end{equation}

Similarly,
\begin{equation}
    2QQ^\dagger-\mathbbm{1}=Q'_{1234}(2|0\rangle_1|0\rangle_3\langle 0|_1\langle 0|_3-\mathbbm{1})Q'^\dagger_{1234}
\end{equation}


Just like in the Quantum SVE, we create a walk operator

\begin{equation}
    W=(2PP^\dagger - \mathbbm{1})(2QQ^\dagger - \mathbbm{1}).
\end{equation}

Then, from Jordan's lemma \cite{jordan}, we can then say that the following states are eigenvectors of $W$ -

\begin{equation}
|\Psi^{(k)}_{i\pm}\rangle_{24}=\frac{Q|\tilde{v_i}^{(k)}\rangle_2|\tilde{k}\rangle_4\mp i(Q|\tilde{v_i}^{(k)}\rangle_2|\tilde{k}\rangle_4)^\perp}{\sqrt{2}}    
\end{equation}

with eigenvalues $\exp{(\pm\theta_i^{(k)})}$.



\begin{figure}[t]
    \centering
    \resizebox{0.9\columnwidth}{!}{
        \begin{quantikz}[row sep=0.5cm,column sep=0.3cm]
            \lstick[1]{\begin{tabular}{@{}c@{}}Left singular \\ vector register\end{tabular}} & \qwbundle{} & \gate[3]{\text{Prepare } |\mathcal{A}\rangle}\slice{$|\psi_1\rangle$} & & & & \\
            \lstick[1]{\begin{tabular}{@{}c@{}}Right singular \\ vector register\end{tabular}} & \qwbundle{} & & \gate[6]{QtSVE}\slice{$|\psi_2\rangle$} & & \gate[6]{QtSVE^\dagger}\slice{$|\psi_4\rangle$} & \\
            \lstick[1]{Face register} & \qwbundle{} & & & & & \\
            \lstick[1]{Walk operator face register} & \qwbundle{} & & & & & \\
            \lstick[1]{Walk operator lsv register} & \qwbundle{} & & & & & \\
            \lstick[1]{Phase register} & \qwbundle{} & & & & & \\
            \lstick[1]{Abs Phase register} & \qwbundle{} & & & \gate[2]{\text{Truncate}}\slice{$|\psi_3\rangle$} & & \\
            \lstick[1]{Flag qubit} & & & & & & \rstick{$|0\rangle$}
        \end{quantikz}
    }
    
    \caption{Circuit diagram of the purely quantum t-SVD truncation algorithm}
    \label{fig:tSVDCircDiag}
\end{figure}
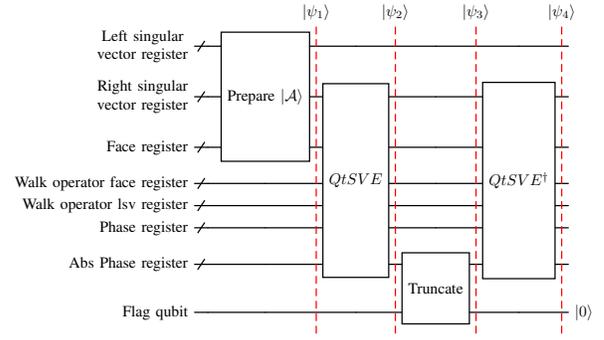

\subsubsection{Algorithm procedure}
Now we will see how we can run the Quantum t-SVE routine on our state $|\tilde{v_i}^{(k)}\rangle|\tilde{k}\rangle$.\\

\textbf{Step 1:} First we apply $Q'$ on $|\tilde{v_i}^{(k)}\rangle|\tilde{k}\rangle$ and two other empty registers to get

\begin{equation}
|\psi_1\rangle=Q'_{1234}|0\rangle_1|\tilde{v_i}^{(k)}\rangle_2|0\rangle_3|\tilde{k}\rangle_4=Q|\tilde{v_i}^{(k)}\rangle|\tilde{k}\rangle.
\end{equation}

This step contributes a circuit depth of:

\begin{equation}
    \text{Circuit depth}=O(\text{polylog}(LMN))
\end{equation}

\textbf{Step 2:} Then we apply Quantum Phase Estimation with respect to $W$ on this state with

\begin{equation}
    n_p=d+\lceil\log_2{(2+\frac{1}{2\epsilon^2})}\rceil.
\end{equation}

This yields us the following state:

\begin{equation}
   |\psi_2\rangle=\frac{|\Psi^{(k)}_{i+}\rangle_{24}|\Theta_i^{(k)}\rangle_{phase}+|\Psi^{(k)}_{i-}\rangle_{24}|-\Theta_i^{(k)}\rangle_{phase}}{\sqrt{2}}
\end{equation}

Here we define $|\Theta_i^{(k)}\rangle$ as follows:

\begin{equation}
    |\Theta_i^{(k)}\rangle=\sqrt{1-\overline{\epsilon}^2}|\theta_i^{(k)}\rangle+\overline{\epsilon}|\dots\rangle
\end{equation}

The state $|\theta_i^{(k)}\rangle$ is defined in the same way that it was in \eqref{eq:binom}.  We're again using $\overline{\epsilon}$ as a placeholder for any value less than or equal to $\epsilon$. We will lean towards overestimating errors, so this quantity won't necessarily remain constant across equations.

This step contributes a circuit depth of:

\begin{align}
    \text{Circuit depth}=&O(\text{polylog}(LMN)\times2^{n_p})\\
    =&O\left(\frac{\text{polylog}(LMN)}{\delta\epsilon^2}\right)
\end{align}

\textbf{Step 3:} Next we copy the absolute value of $\theta_i^{(k)}$ to a new register and uncompute the QPE step. Now we get the state

\begin{equation}
    |\psi_3\rangle=Q'_{1234}|0\rangle_1|\tilde{v_i}^{(k)}\rangle_2|0\rangle_3|\tilde{k}\rangle_4|\tilde{\theta}_i^{(k)}\rangle_{abs}+\sqrt{2}\overline{\epsilon}|\dots\rangle,
\end{equation}

where $|\tilde{\theta}_i^{(k)}\rangle$ represents the basis state approximation of $2^d\times\theta_i^{(k)}/(2\pi)$.

To do this, we need an operator $K$ such that

\begin{equation}
    K|x\rangle|0\rangle=|x\rangle||x|\rangle
\end{equation}

To implement this, we can first use Toffoli gates to copy the first $d$ qubits of the phase register to the absolute phase register controlled for when the most significant qubit is $|0\rangle$ ($O(n_p)$), i.e., when the phase register is in some basis state $|x\rangle$ where $x\geq0$. Then we can use the QFT weighted adder from \cite{qftadder} to set the absolute phase register to $|-x\rangle$ controlled on when the most significant qubit is in the $|1\rangle$ state ($O(n_p^2)$). The circuit depth of $K$ is thus $O(n_p^2)$. This depth is insignificant compared to uncomputing the QPE step, so the time complexity of this step again comes out to be

\begin{align}
    \text{Circuit depth}=&O(\text{polylog}(LMN)\times2^{n_p})\\
    =&O\left(\frac{\text{polylog}(LMN)}{\delta\epsilon^2}\right).
\end{align}


\textbf{Step 4:} Applying $Q'^\dagger$, we finally get the output state we wanted:


\begin{align}
    |\psi_4\rangle=&|\tilde{v_i}^{(k)}\rangle_2|\tilde{k}\rangle_4|\tilde{\theta}_i^{(k)}\rangle_{abs}+\sqrt{2}\overline{\epsilon}|\dots\rangle
\end{align}

This step contributes to a circuit depth of

\begin{equation}
     \text{Circuit depth}=O(\text{polylog}(LMN))
\end{equation}

We will represent this whole procedure as an operator denoted by $QtSVE_{1234}$. The leading contribution to the circuit depth comes from the QPE step, and so the time complexity of this operation is given by

\begin{equation}
    \text{Circuit depth}=O\left(\frac{\text{polylog}(LMN)}{\delta\epsilon^2}\right).
\end{equation}

\subsection{Tensor truncation with t-SVD}\label{subsec:tensortrunc}







We define $\mathcal{A}$, $\theta^{(k)}_i$, $\delta$, $d$, $\epsilon$ and $\overline{\epsilon}$ as they were defined in Section \ref{sec:qtsvedefs}
Let $\mathcal{A}^{\tau}_{trunc}$ be the truncated tensor obtained by setting all the singular values of $\mathcal{A}$ less than $\tau$ to $0$.

The output of this algorithm will be the state

\begin{equation}
    |\mathcal{A}^{\tau}_{trunc}\rangle + {2}\overline{\epsilon}|\dots\rangle
\end{equation}

with the probability of success being $\alpha^2$. We define $\alpha$ as

\begin{equation}
    \alpha:=\frac{\|\mathcal{A}^{\tau}_{trunc}\|_F}{\|\mathcal{A}\|_F}
\end{equation}


\subsubsection{Circuit registers}
This circuit is depicted in Fig. \ref{fig:tSVDCircDiag} and makes use of $8$ registers:
\begin{enumerate}[(i)]
    \item Left singular vector register $(1)$, with $n_m:=\lceil \log_2{(M)}\rceil$ qubits
    \item Right singular vector register $(2)$, with $n_n:=\lceil \log_2{(N)}\rceil$ qubits
    \item Face register $(3)$, with $n_l:=\lceil \log_2{(L)}\rceil$ qubits
    \item Walk operator face register $(4)$, with $n_l$ qubits
    \item Walk operator left singular vector register 2 $(5)$, with $n_m$ qubits
    \item Phase register $(6)$, with $n_p$ qubits
    \item Abs Phase register $(7)$, with $d$ qubits
    \item Flag qubit $(F)$
\end{enumerate}

\subsubsection{Algorithm procedure}
We begin with all qubits initialized to the $|0\rangle$ state. The procedure is as follows.

\textbf{Step 1:} Prepare the state

\begin{align}
|\psi_1\rangle=&\frac{1}{\|\mathcal{A}\|_F}\sum_{ijk}a_{ijk}|i\rangle_1|j\rangle_2|k\rangle_3\\
=&\frac{1}{\|\mathcal{A}\|_F}\sum_{ijk}\tilde{a}_{ijk}|i\rangle_1|j\rangle_2|\tilde{k}\rangle_3\nonumber\\
=&\frac{1}{\|\mathcal{A}\|_F}\sum_{tk}\tilde{\sigma}^{(k)}_t|\tilde{u}^{(k)}_t\rangle_1|\tilde{v}^{(k)}_t\rangle_2|\tilde{k}\rangle_3.
\end{align}

To get this state, we start with $|0\rangle_1|0\rangle_2|0\rangle_3|0\rangle_4$ and apply Hadamard gates on all the qubits in register $3$ to get $|0\rangle_1|0\rangle_2|\tilde{0}\rangle_3|0\rangle_4$. Then we apply $P'_{1234}$ to get

\begin{align}
    &P'_{1234}|0\rangle_1|0\rangle_2|\tilde{0}\rangle_3|0\rangle_4\nonumber\\
    =&\frac{1}{\|\mathcal{A}\|_F}\sum_j\|\mathcal{A}[:,j,:]\|_F|0\rangle_1|j\rangle_2|\tilde{0}\rangle_3|\tilde{0}\rangle_4
\end{align}

\begin{figure}[htbp]
    \centering
    \begin{subfigure}[b]{0.33\textwidth}
        \centering
        \includegraphics[width=\textwidth]{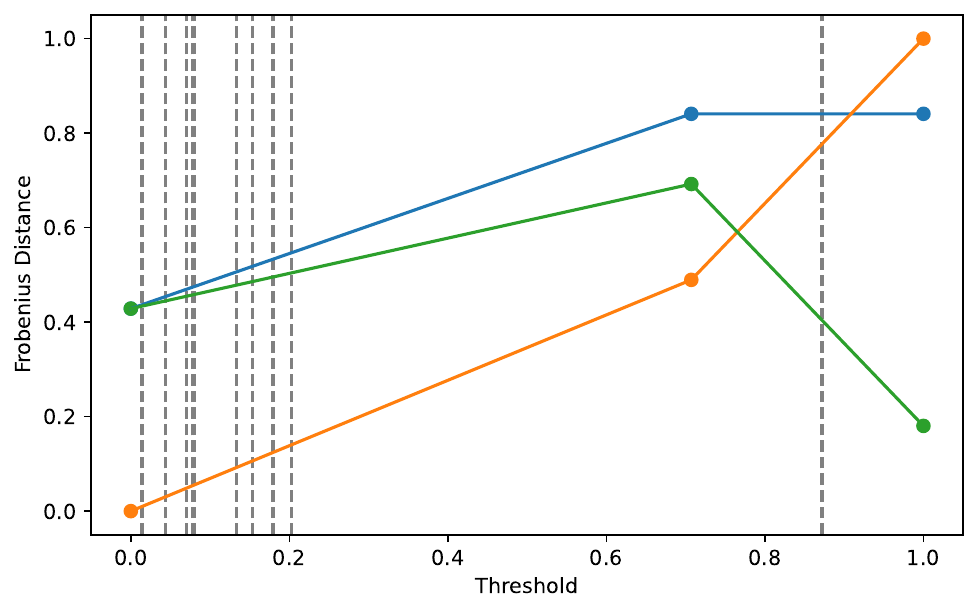}
        \caption{2 phase qubits}
    \end{subfigure}\\
    \vspace{0.5cm}
    \begin{subfigure}[b]{0.33\textwidth}
        \centering
        \includegraphics[width=\textwidth]{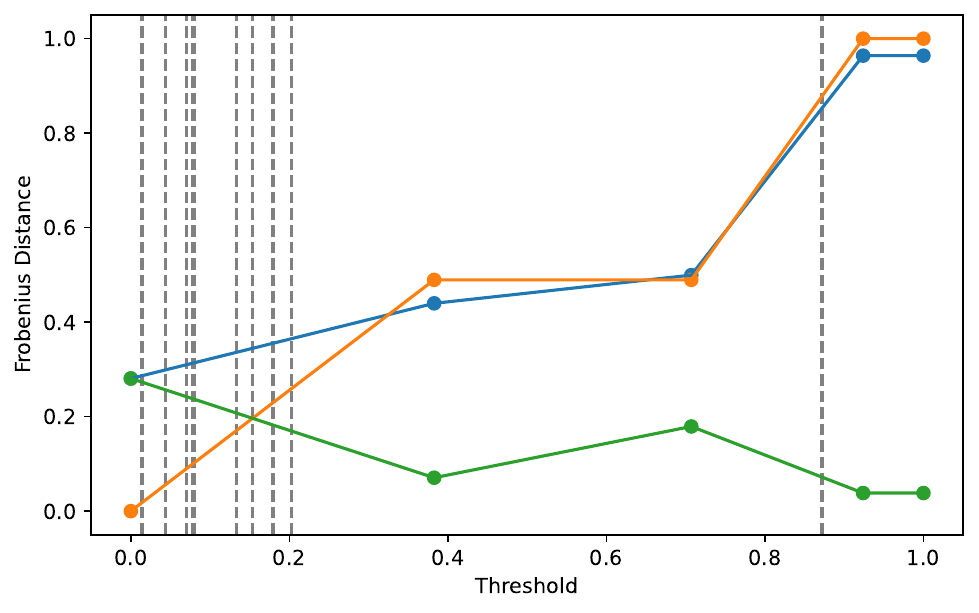}
        \caption{3 phase qubits}
    \end{subfigure}
    \\
    \vspace{0.5cm}
    \begin{subfigure}[b]{0.33\textwidth}
        \centering
        \includegraphics[width=\textwidth]{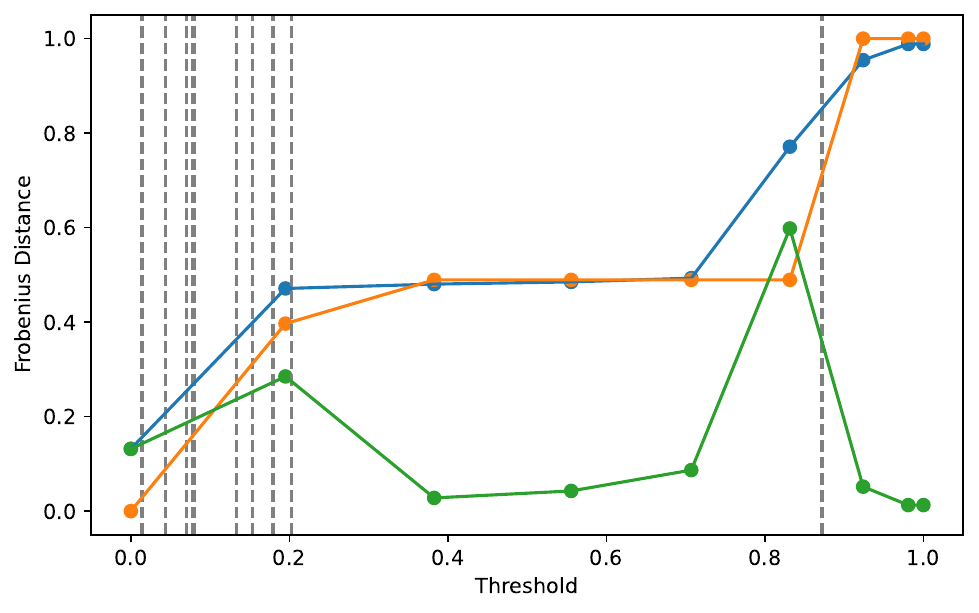}
        \caption{4 phase qubits}
    \end{subfigure}\\
    \vspace{0.5cm}
    \begin{subfigure}[b]{0.33\textwidth}
        \centering
        \includegraphics[width=\textwidth]{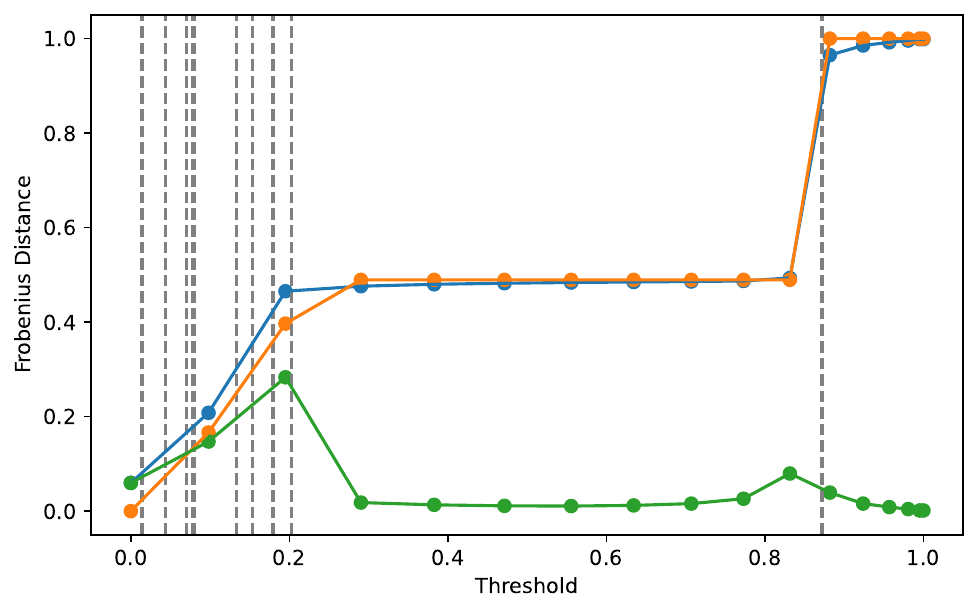}
        \caption{5 phase qubits}
    \end{subfigure}\\
    \vspace{0.5cm}
    \hspace*{-2.3cm}\begin{tikzpicture}
    \definecolor{darkorange25512714}{RGB}{255,127,14}
    \definecolor{forestgreen4416044}{RGB}{44,160,44}
    \definecolor{steelblue31119180}{RGB}{31,119,180}
        \begin{axis}[%
        hide axis,
        xmin=10,
        xmax=50,
        ymin=0,
        ymax=0.4,
        legend style={draw=white!15!black,legend cell align=left}
        ]
        \addlegendimage{steelblue31119180}
        \addlegendentry{$\|\mathcal{A}^{q}_{trunc}-\mathcal{A}\|_F$};
        \addlegendimage{darkorange25512714}
        \addlegendentry{$\|\mathcal{A}^{c}_{trunc}-\mathcal{A}\|_F$};        
        \addlegendimage{forestgreen4416044}
        \addlegendentry{$\|\mathcal{A}^{q}_{trunc}-\mathcal{A}^{c}_{trunc}\|_F$};
        \end{axis}        
    \end{tikzpicture}
    \vspace{-105pt}\\
        \footnotesize{Legend}
    \vspace{10pt}
    \caption[Simulating the t-SVD tensor completion algorithm for a $4\times4\times4$ tensor]{Simulating the t-SVD tensor completion algorithm for a $4\times4\times4$ tensor whose each entry is randomly sampled from a uniform distribution. The $x$-axis represents different choices of threshold $\tau$. The grey dashed lines represent the t-SVD singular values of the tensor.}
    \label{fig:tSVDres}
\end{figure}

Next, we apply Hadamard gates on registers $3$ and $4$ to reset those registers to $|0\rangle$ and then apply $Q'_{1234}$ to finally get

\begin{equation}
    |\psi_1\rangle=\frac{1}{\|\mathcal{A}\|_F}\sum_{ijk}a_{ijk}|i\rangle_1|j\rangle_2|k\rangle_3|0\rangle_4.
\end{equation}

This step contributes a circuit depth of:

\begin{equation}
    \text{Circuit depth}=\mathcal{O}(\text{polylog}(LMN))
\end{equation}

For \textbf{Step 2:} Apply the operator $QtSVE_{5243}$ to store $|\theta^{(k)}_t\rangle$ in a new register.

\begin{align}
|\psi_2\rangle=&\frac{1}{\|\mathcal{A}\|_F}\sum_{tk}\tilde{\sigma}^{(k)}_t|\tilde{u}^{(k)}_t\rangle_1|\tilde{v}^{(k)}_t\rangle_2|\tilde{k}\rangle_3|\tilde{\theta}^{(k)}_t\rangle_7+\sqrt{2}\overline{\epsilon}|\dots\rangle
\end{align}


This step contributes a circuit depth of:

\begin{equation}
    \text{Circuit depth}=O\left(\frac{\text{polylog}(LMN)}{\delta\epsilon^2}\right).
\end{equation}





\textbf{Step 3:} Use a flag qubit to separate the singular values above a given threshold $\tau$.




\begin{align}
|\psi_3\rangle=&\frac{1}{\|\mathcal{A}\|_F}\sum_{{\sigma}^{(k)}_t>\tau}\tilde{\sigma}^{(k)}_t|\tilde{u}^{(k)}_t\rangle_1|\tilde{v}^{(k)}_t\rangle_2|\tilde{k}\rangle_3|\tilde{\theta}^{(k)}_t\rangle_7|0\rangle_F\nonumber\\
&+\sqrt{1-\alpha^2}|\dots\rangle|1\rangle_F+\sqrt{2}\overline{\epsilon}|\dots\rangle
\end{align}

For this, we need to construct an operator $G$ such that

\begin{equation}
    G|\theta\rangle_7|0\rangle_F=|\theta\rangle_7|f_{\tau/\|\mathcal{A}\|_F}(\theta)\rangle_F.
\end{equation}

where we define
\begin{equation}
    f_a(\theta):=\begin{cases} 
      1 & \cos{\frac{\theta}{2}}<a \\
      0 & \cos{\frac{\theta}{2}}\geq a
    \end{cases}
\end{equation}

Since $\theta\in[0,\pi]$, the condition $\cos{\frac{\theta}{2}}<a$ is equivalent to $\theta>2\arccos{a}$.  This can be implemented using a simple integer comparator \cite{intcomp} on the first $d$ qubits of the absolute phase register. This circuit has time complexity:

\begin{equation}
    \text{Circuit depth}=O\left(\frac{1}{\delta}\right)
\end{equation}

\textbf{Step 4:} Apply the inverse of the $QtSVE$ operator to get the output state:


\begin{align}
    |\psi_4\rangle=&\alpha|\mathcal{A}^\tau_{trunc}\rangle|0\rangle_F+\sqrt{1-\alpha^2}|\dots\rangle|1\rangle_F+2\overline{\epsilon}|\dots\rangle\label{eq:truncout}
\end{align}







Then we postselect for the flag qubit to be in the $|0\rangle_F$ state. The minimum probability of measuring this state can be calculated as:

\begin{align}
    \text{Probability of success} > (\alpha-2\epsilon)^2
\end{align}

We now write down the form of the final output state after postselection, while making worst-case assumptions regarding the error terms:



\begin{equation}
    |\psi\rangle\approx|\mathcal{A}^\tau_{trunc}\rangle+ {2}\overline{\epsilon}|\dots\rangle.
\end{equation}

Time complexity:

\begin{equation}
    \text{Circuit depth}=O\left(\frac{\text{polylog}(LMN)}{\delta\epsilon^2}\right).
\end{equation}

\subsection{Quantum t-SVD Recommendation System}
We may turn this into a quantum recommendation system that can take in the index of an input entry, $i$, and recommend a primary output index, $j$, along with an associated value, $k$. For example, a potential use-case would be if a store wanted to email discount codes for a certain product ($i$) to a few customers. They would have to decide which customer ($j$) to target this discount to, and on what day of the week ($k$) the customer should be sent these emails (based on when the customer tends to do their shopping) in order to maximize incentive. To do so, we employ tensor truncation as a method of tensor completion (similar to \cite{tsvd}, and to how \cite{kp} uses matrix truncation as a method of matrix completion). Given an incomplete preference tensor $\mathcal{A}$ whose unknown elements are set to $0$, we assume our data is structured such that a t-SVD truncation of this incomplete tensor approximates the true complete preference tensor $\mathcal{A}^\tau_{trunc}$.\\

To run this algorithm, we need to follow the same procedure as in Section \ref{subsec:tensortrunc} except we change the initial state in Step 1. Here, we start by preparing the basis state $|i\rangle_1$ as input. Then we apply an operator $R$ such that

\begin{equation}
    R|i\rangle_1|0\rangle_2|0\rangle_3 = \frac{1}{\|\mathcal{A}[i,:,:]\|_F}\sum_{jk}a_{ijk}|i\rangle_1|j\rangle_2|k\rangle_3
\end{equation}

One way such an operator can be constructed is by following the same method with which $Q'$ was constructed, using an additional data structure to store a matrix whose $(i,j\times 2^{n_l}+k)$-th entry is $\mathcal{A}[i,j,k]$.

After following the rest of the procedure from Section \ref{subsec:tensortrunc}, we postselect for $|0\rangle_F$ and get the following state:

\begin{equation}
    |\psi\rangle\approx|i\rangle|\mathcal{A}^\tau_{trunc}[i,:,:]\rangle+ {2}\overline{\epsilon}|\dots\rangle
\end{equation}

We can then make a measurement on registers $2$ and $3$ to get our recommendations $j$ and $k$ respectively. Since we only need to make one measurement shot to get a recommendation, the time complexity of this algorithm is also just the circuit depth:

\begin{equation}
    \text{Circuit depth}=O\left(\frac{\text{polylog}(LMN)}{\delta\epsilon^2}\right).
\end{equation}

\subsection{Running simulations}

We use QuEST \cite{quest} to simulate the truncation algorithm for a randomly generated $4\times4\times4$ tensor (each entry sampled from a uniform distribution) which is then scaled to have a Frobenius norm of $1$. These simulations are run for various values of $\tau$ and $n_p$ and the results are depicted in Fig. \ref{fig:tSVDres}.

The results are compared with classical methods of truncation by plotting the Frobenius distances between (i) The original tensor $\mathcal{A}$ and the tensor generated through the simulated quantum algorithm $\mathcal{A}^{q}_{trunc}$ (blue), (ii) The original tensor $\mathcal{A}$ and the tensor generated through a classical truncation procedure $\mathcal{A}^{c}_{trunc}$ (orange), and (iii) The tensor generated through the simulated quantum algorithm $\mathcal{A}^{q}_{trunc}$ and the one generated through classical truncation $\mathcal{A}^{c}_{trunc}$ (green). These Frobenius distances are plotted against the threshold $\tau$ on the x-axis. From plots (i) and (ii), we see that the truncated tensors $\mathcal{A}^{q}_{trunc}$ and $\mathcal{A}^{q}_{trunc}$ stray further from the original tensor $\mathcal{A}$ in Frobenius distance as the threshold increases and more singular values are set to $0$. As we can expect, we see that the quantum results in plot (i) fit more tightly onto the classically calculated results in plot (ii) as we increase the number of phase register qubits. The difference between their results is measured in plot (iii), and we can see that these differences show up around the singular values. The classical computation switches sharply when the threshold crosses a singular value, whereas the quantum computation interpolates smoothly. We therefore see spikes in (iii) around such points.



\begin{figure}[t]
    \centering
  \includegraphics[width=0.8\columnwidth]{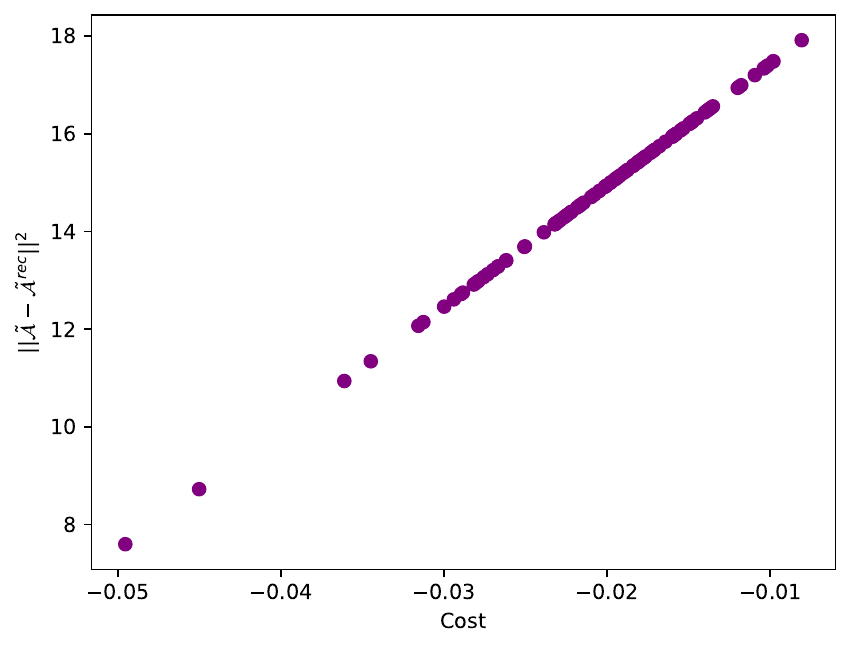}
 \caption[Error vs Cost for Variational t-SVD]{Error vs Cost of $100$ instances of randomly sampled parameters for a fixed randomly generated $4\times4\times4$ tensor $\mathcal{A}$ with $T=4$}
 \label{fig:errvscostvtsvd}
 \end{figure}

\begin{figure}[t]
   \centering
 \includegraphics[width=0.8\columnwidth]{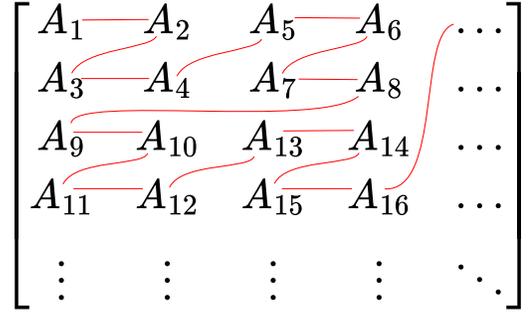}
\caption{Ordering of $A_{flat}$}
\label{fig:zorder}
\end{figure}

\section{Variational algorithm for \normalfont{t}-SVD}\label{sec:VtSVD}
In this section we describe a new variational quantum t-SVD routine constructed similarly to \cite{moi}.

\subsection{Broad overview}\label{sec:vtsvdoverview}

The algorithm takes the following inputs
\begin{enumerate}
    \item A $2^n\times2^n\times2^k$ tensor $\mathcal{A}$
    \item A positive integer $1\leq T\leq 2^n$
\end{enumerate}

If the tensor we want the t-SVD of is not $2^n\times2^n\times2^k$, we can always pad it with zeros to make it so.

The algorithm involves a parameterized ansatz $W$ and two sets of parameters $\Vec{\alpha}$ and $\Vec{\beta}$ that are optimized with respect to an objective function.
$W$ is a special ansatz in that it takes in two registers $1$ and $2$ of size $k$ and $n$ respectively and leaves computational basis states in register $1$ unaffected. That is to say,
\begin{equation}
    W|m\rangle_1|\phi\rangle_2=|m\rangle_1|\psi\rangle_2,
\end{equation}

for some $\psi$ and $\phi$. Another way of phrasing this is that $W$ has a block diagonal matrix representation, where each $2^n\times2^n$ block is given by $\langle m|_1 W |m \rangle_1$. Define $U'_m$ and $V'_m$ as

\begin{align}
    U'_m:=&\langle m|_1 W(\vec{\alpha}) |m \rangle_1\\
    V^{\prime\dagger}_m:=&\langle m|_1 W(\vec{\beta}) |m \rangle_1
\end{align}

The algorithm attempts to find unitaries $\{U^{\prime\dagger}_m\}$ and $\{V'_m\}$ that maximize the following expression:

\begin{equation}
    \frac{1}{\|\mathcal{A}\|_F^2 2^n}\sum_m\sum_i|\langle i|U^{\prime\dagger}_m\tilde{A}^{(m)}V'_m|i\rangle|^2\label{eq:tsvdobj}
\end{equation}

Since these inner sums are (ideally, for an expressive ansatz) independent of each other, maximizing the whole expression is equivalent to maximizing all the inner sums. From \cite{wang} we know that when the inner sum is maximized,

\begin{align}
    \tilde{\mathcal{A}}^{rec}[:,:,m]:=&\sum_{i<T} \Tilde{\sigma}^{\prime (m)}_i U'_m|i\rangle\langle i|V^{\prime\dagger}_m\nonumber\\
    \approx&\sum_{i<T} \Tilde{\sigma}^{(m)}_i U^{(m)}|i\rangle\langle i|V^{(m)\dagger}.\label{eq:vtsvdapprox}
\end{align}

Here,
\begin{equation}
    {(\Tilde{\sigma}^{\prime (m)}_i)}^2:=|\langle i|U^{\prime\dagger}_m\tilde{A}^{(m)}V'_m|i\rangle|^2
\end{equation}
and $\Tilde{\sigma}^{(m)}_i$ represents the $i$-th largest singular value of $\tilde{A}^{(m)}$.\\
This can be seen in Fig. \ref{fig:errvscostvtsvd}, which plots $\|\tilde{\mathcal{A}}-\tilde{\mathcal{A}}^{rec}\|_F$ against the cost (negative of the objective function) of $100$ instances of $\tilde{\mathcal{A}}^{rec}$ produced with randomly sampled parameters $\vec{\alpha}$ and $\vec{\beta}$ for a randomly generated $4\times4\times4$ tensor $\tilde{\mathcal{A}}$ with $T=4$. As we can see from the figure, a lower cost corresponds directly to a lower error.


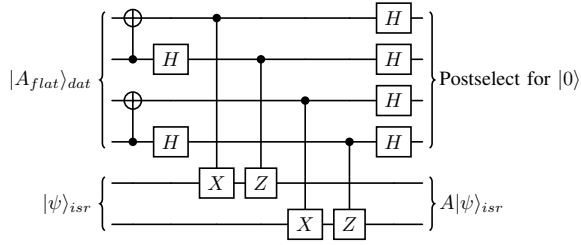
\begin {figure}[t]
\centering
\resizebox{0.9\columnwidth}{!}{
\begin{quantikz}[row sep=0.2cm,column sep=0.2cm]
\\
 \lstick[4]{$|A_{flat}\rangle_{dat}$} & \targ{} & & \ctrl{4} & & & & \gate{H} & \rstick[4]{Postselect for $|0\rangle$} \\
 & \ctrl{-1} & \gate{H} & & \ctrl{3} & & & \gate{H} & \\
 & \targ{} & & & & \ctrl{3} & & \gate{H} & \\
 & \ctrl{-1} & \gate{H} & & & & \ctrl{2} & \gate{H} & \\
 \lstick[2]{$|\psi\rangle_{isr}$} & & & \gate{X} & \gate{Z} & & &  & \rstick[2]{$A|\psi\rangle_{isr}$}\\
 & & & & & \gate{X} & \gate{Z} & &
\end{quantikz}
}
\caption{Full block encoding circuit for a $4\times 4$ matrix $A$ using the new method.}
\label{fig:novelbe}
\end{figure}

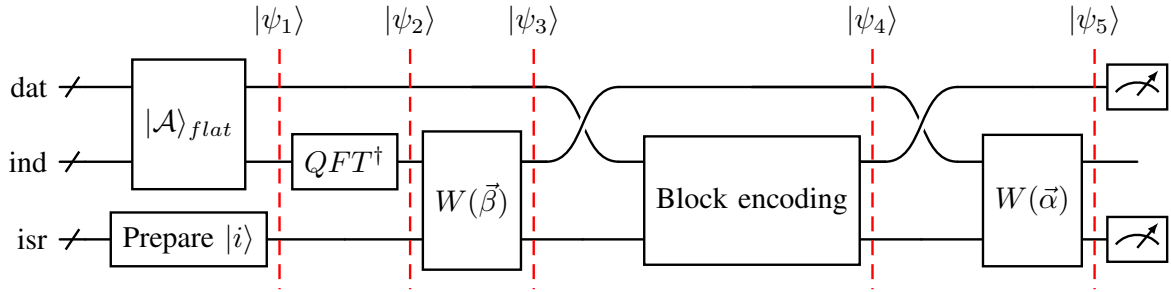
\begin{figure*}
    \centering
    \resizebox{1.8\columnwidth}{!}{
        \begin{quantikz}[row sep=0.2cm,column sep=0.3cm]
            \lstick[1]{dat} & \qwbundle{} & \gate[2]{ |\mathcal{A}\rangle_{flat} }\slice{$|\psi_1\rangle$} & & & \permute{2,1} & & \permute{2,1} & & \meter{} \\
            \lstick[1]{ind} & \qwbundle{} & & \gate{QFT^\dagger}\slice{$|\psi_2\rangle$} & \gate[2]{W(\vec{\beta})}\slice{$|\psi_3\rangle$} &  & \gate[2]{\text{Block encoding}}\slice{$|\psi_4\rangle$} & & \gate[2]{W(\vec{\alpha})}\slice{$|\psi_5\rangle$} & \\
            \lstick[1]{isr} & \qwbundle{} & \gate{\text{Prepare } |i\rangle} & & & & & & & \meter{}
        \end{quantikz}
    }
    
    \caption{Circuit diagram of the variational quantum t-SVD algorithm}
    \label{fig:VtSVDCircDiag}
\end{figure*}

\subsection{Block encoding with quantum data}\label{subsec:novelbe}

Variational quantum algorithms work by using parameterized quantum circuits to evaluate the given objective function. However, the terms in \eqref{eq:tsvdobj} are not easy to estimate. One reason for this is that $\tilde{A}^{(m)}$ is not unitary, and so $U^{\prime\dagger}_m\tilde{A}^{(m)}V'_m$ is not a valid operation we can take the expectation value of. The solution to this is to use a block encoding of $\tilde{A}^{(m)}$. But this brings us to the second issue; we don't have classical knowledge of $\tilde{A}$ and most block encodings rely on classical information of the matrix that needs to be encoded. We could classically compute it, but this would seem wasteful given that quantum computers are good at taking Fourier transforms.

Our solution is to use our block encoding method proposed in \cite{moi}. Using this method we can construct a circuit (depicted in Fig. \ref{fig:novelbe}) which takes an amplitude encoding of $A$ in one register and applies $A$ as an operator on a second register when we postselect the first register to be in the $|0\rangle$ state. Amplitude encoding $A$ involves flattening it into a statevector using the ordering depicted in Fig. \ref{fig:zorder}.

\begin{figure}[htbp]
    \centering
  \includegraphics[width=0.8\columnwidth]{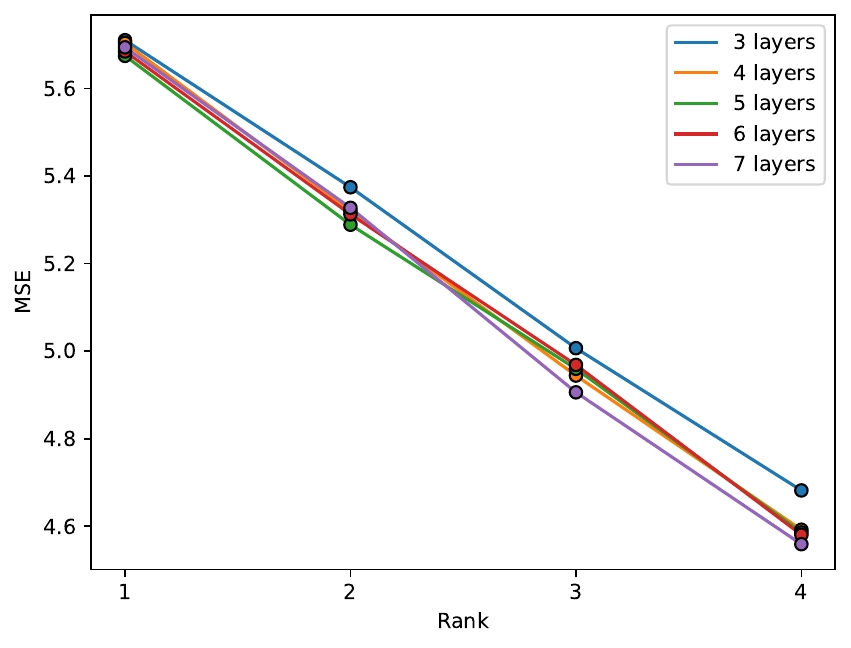}
 \caption{MSE vs T when simulating Variational t-SVD with different ansatz layer counts}
 \label{fig:vtsvdres}
 \end{figure}

\subsection{Algorithm procedure}

In this section, the algorithm is laid out in detail.

We start with a description of the procedure to obtain the objective function in \eqref{eq:tsvdobj} using quantum circuits.

The quantum circuit we use is depicted in Fig. \ref{fig:VtSVDCircDiag} and has $3$ registers:

\begin{enumerate}[(i)]
    \item Face index register with $l$ qubits ($ind$)
    \item Face data register with $2n$ qubits ($dat$)
    \item Input state register with $n$ qubits ($isr$)
\end{enumerate}

The procedure to obtain the state

\begin{equation}
    |\psi\rangle=\frac{1}{\|\mathcal{A}\|_F 2^n}\sum_m|\langle i|U^{\prime\dagger}_m\tilde{A}^{(m)}V'_m|i\rangle|^2
\end{equation}

for a given $i$ is as follows:

\textbf{Step 1a:} Prepare $|i\rangle$ on the input state register.\\

\textbf{Step 1b:} Let $\mathcal{A}_{flat}$ represent the normalized vector obtained by concatenating the Z-order flattening of each face $A^{(l)}_{flat}$ one after the other. We prepare this statevector in the face index register and face data register to get the state:

\begin{equation}
    |\psi_1\rangle=|\mathcal{A}_{flat}\rangle_{ind,dat}|i\rangle_{isr} = \sum_k^{2^l} |k\rangle_{ind}|A^{(k)}_{flat}\rangle_{dat} |i\rangle_{isr}
\end{equation}

\begin {figure*}[t]
\centering
\begin{quantikz}[row sep=0.2cm,column sep=0.2cm]
\lstick[2]{face\\index}  &          &          &           &           & \ctrl{2}  & \ctrl{3}  & \ctrl{4}  &           &           &           & \ctrl{2}  & \ctrl{3}  & \ctrl{4}  &           &           &           &\\
                            &          &          &           &           &           &           &           & \ctrl{1}  & \ctrl{2}  & \ctrl{3}  &           &           &           & \ctrl{1}  & \ctrl{2}  & \ctrl{3}  &\\
\lstick[3]{input\\state} & \ctrl{1} &          & \gate{RZ} & \gate{RY} & \gate{RZ} &           &           & \gate{RZ} &           &           & \gate{RY} &           &           & \gate{RY} &           &           &\\
                            & \targ{}  & \ctrl{1} & \gate{RZ} & \gate{RY} &           & \gate{RZ} &           &           & \gate{RZ} &           &           & \gate{RY} &           &           & \gate{RY} &           &\\
                            &          & \targ{}  & \gate{RZ} & \gate{RY} &           &           & \gate{RZ} &           &           & \gate{RZ} &           &           & \gate{RY} &           &           & \gate{RY} &\\
\end{quantikz}
\caption{A single layer of the ansatz used in simulating the Variational t-SVD}
\label{fig:vtsvdansatz}
\vspace{-5mm}
\end{figure*}
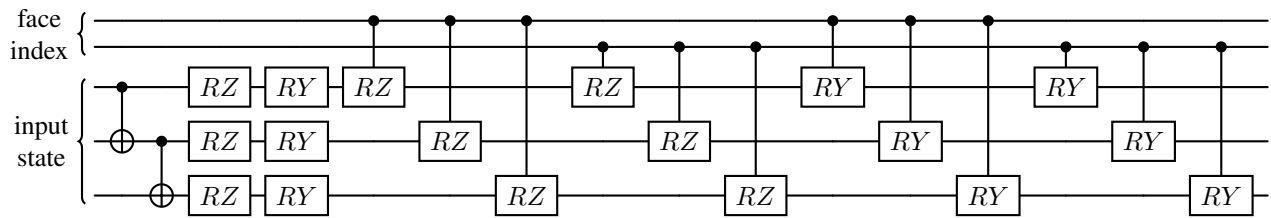

\textbf{Step 2:} Apply an inverse Quantum Fourier Transform on the face index register to obtain the state:

\begin{equation}
    |\psi_2\rangle=\sum_m^{2^l} |m\rangle_{ind}|\tilde{A}^{(m)}_{flat}\rangle_{dat}|i\rangle_{isr}
\end{equation}

\textbf{Step 3:} We apply ansatz $W(\vec{\beta})$ on the face index register and the input state register to get the following state:

\begin{equation}
    |\psi_3\rangle=\sum_m^{2^l} |m\rangle_{ind}|\tilde{A}^{(m)}_{flat}\rangle_{dat}V'_m|i\rangle_{isr}
\end{equation}

\textbf{Step 4:} We apply the block encoding from Section \ref{subsec:novelbe} on the face data register and input state register to produce the state:

\begin{align}
    |\psi_4\rangle=&\frac{1}{\|\mathcal{A}\|_F\sqrt{2^n}}\sum_m^{2^l} |m\rangle_{ind}|0\rangle_{dat} \tilde{A}^{(m)}V'_m|i\rangle_{isr}\nonumber\\
    &+ |0^\perp\rangle_{dat}|\dots\rangle_{ind,isr}
\end{align}

\textbf{Step 5:} We apply ansatz $W^\dagger(\vec{\alpha})$ on the face index register and the input state register to get the following state:
\begin{align}
    |\psi_5\rangle=&\frac{1}{\|\mathcal{A}\|_F\sqrt{2^n}}\sum_m^{2^l} |m\rangle_{ind}|0\rangle_{dat} U^{\prime\dagger}_m\tilde{A}^{(m)}V'_m|i\rangle\nonumber\\
    &+ |0^\perp\rangle_{dat}|\dots\rangle_{ind,isr}\label{eq:vtsvdfinalstate}
\end{align}

\textbf{Step 6:} Estimate the probability of measuring $|0\rangle_{dat}|i\rangle_{isr}$. Born's rule dictates that this probability will be equal to
\begin{equation}
\text{Pr}(|0\rangle_{dat}|i\rangle_{isr})=\frac{1}{\|\mathcal{A}\|_F^2 {2^n}}\sum_m|\langle i|U^{\prime\dagger}_m\tilde{A}^{(m)}V'_m|i\rangle|^2, \label{eq:vtsvdeval}
\end{equation}
which is the required quantity.

The full objective function \eqref{eq:tsvdobj} can be evaluated by running this circuit for all $0\leq i<T$ and summing all the results \eqref{eq:vtsvdeval}. 

We use a classical optimizer to find the optimal parameters $\vec{\alpha}^*$ and $\vec{\beta}^*$ that maximize this objective function. As mentioned in Section \ref{sec:vtsvdoverview}, we then get $U^{\prime\dagger}_m\approx U^{(m)\dagger}$, $V'_m\approx V^{(m)}$, and $|\langle i|U^{\prime\dagger}_m\tilde{A}^{(m)}V'_m|i\rangle|^2\approx |\tilde{\sigma}^{(m)}_i|^2$. The latter can be obtained by running the circuit for optimal parameters and estimating the probability of measuring $|m\rangle_{ind}|0\rangle_{dat}|i\rangle_{isr}$ from state \eqref{eq:vtsvdfinalstate}. This probability comes out to be

\begin{align}
    \text{Pr}(|m\rangle_{ind}|0\rangle_{dat}|i\rangle_{isr})&=\frac{1}{\|\mathcal{A}\|_F^2 2^n}|\langle i|U^{\prime\dagger}_m\tilde{A}^{(m)}V'_m|i\rangle|^2\nonumber\\
    &\approx |\tilde{\sigma}^{(m)}_i|^2.
\end{align}

\subsection{Running simulations}
This algorithm is implemented in PennyLane \cite{pennylane} and simulations have been run on $60$ randomly generated $4\times4\times4$ tensors whose entries have been sampled from a uniform distribution.

The algorithm is run for each tensor $\mathcal{A}$, and the output tensor $\mathcal{A}^{rec}$ is produced, as defined in \eqref{eq:vtsvdapprox}, for different values of $T$. Error is measured as the Frobenius distance between $\mathcal{A}^{rec}$ and $\mathcal{A}$. The MSE is plotted for different values of $T$ and ansatz layer counts in Fig. \ref{fig:vtsvdres}. We observe that the MSE decreases with rank, as expected. We also observe however, that the performance of the algorithm does not seem to change in any significant way as we increase the number of layers in our ansatz.  The ansatz we used is depicted in Fig. \ref{fig:vtsvdansatz}.

\section{Discussion}\label{sec:disc}
In this work, we have introduced two new quantum algorithms for t-SVD. The first is a purely quantum algorithm and we have shown that it has a runtime that is polylogarithmic in the size of the tensor. This is exponentially faster than currently known classical t-SVD algorithms. For future works, dequantization efforts need to be explored to see if our algorithm truly is better than what can be done with classical computing. We also need to identify useful applications for both algorithms and benchmark these algorithms for such real use-cases. The second algorithm is a variational approach that makes use of our new block encoding from \cite{moi}. This algorithm showcases the utility of being able to apply any operator that's encoded into the amplitudes of a statevector. This block encoding method translates our ability to manipulate the amplitudes of a statevector, into manipulations of operators. The simulation results however indicate poor scaling in performance as the layer count of the ansatz increases. Different choices of ans\"{a}tze and higher layer counts should be explored. In both algorithms, the QFT could be replaced by other transformations, like the Hadamard transform for example. It is worth exploring which transformation would lead to the best results for a given application.

\bibliographystyle{IEEEtran}
\bibliography{IEEEabrv,citations}

\end{document}